\documentstyle[preprint,eqsecnum,aps]{revtex}

\begin{document}
\draft
\title{Real-axis direct solution of the $d$-wave Eliashberg equations and the
tunneling density of states in optimally doped Bi$_2$Sr$_2$CaCu$_2$O$_{8+x}$}
\author{G.A. Ummarino and R.S. Gonnelli}
\address{$INFM-$Dipartimento di Fisica, Politecnico di Torino, 10129 Torino, Italy}
\maketitle

\begin{abstract}
In this work we calculate the direct solution of the equations for the
retarded electron-boson interaction in the case of {\it d}-wave symmetry for
the pair wave function and in the real axis formulation. We use a spectral
function containing an isotropic part and an anisotropic one: $\alpha
^2(\Omega ,\phi ,\phi ^{\prime })F(\Omega )=\alpha _s^2F(\Omega )+\alpha
_d^2F(\Omega )\cdot \cos \left( 2\phi \right) \cos \left( 2\phi ^{\prime
}\right) $ and make the simple assumption: $\alpha _d^2F(\Omega )=g_d\alpha
_s^2F(\Omega )$ where $g_d$ is a constant. For appropriate values of the
isotropic electron-boson coupling constant $\lambda _s$ and the anisotropic
one $\lambda _d$, solutions are obtained with only {\it d}-wave symmetry for
the order parameter and only {\it s}-wave one for the renormalization
function. We have employed the real axis formulation in order to compare the
theoretical curves to the tunneling density of states of the optimally-doped
high-$T_c$ superconductor Bi$_2$Sr$_2$CaCu$_2$O$_{8+x}$. The results of our
numerical simulations are able to fit very well the value of the gap, the
critical temperature and the shape of the density of states in all the
energy range, as recently determined in our break-junction tunneling
experiments. At $T>T_c$ the theoretical conductance still shows a broaden
peak that disappears at $T^{*}\approx 140$ K.
\end{abstract}


\preprint{HEP/123-qed}

\narrowtext

At the present moment, there are many experimental results that strongly
suggest the presence of a dominant {\it d}-wave symmetry of the pair wave
function in high-$T_c$ cuprates, at least the layered ones \cite{refe0}.
Recent tunneling measurements on the layered high-$T_c$ superconductor Bi$_2$%
Sr$_2$CaCu$_2$O$_{8+x}$ (Bi 2212) \cite
{refe4,refe5,refe6,refe8,refe10,refe11,refe12} yielded high or very high
values of the low-temperature gap, gapless features and a small but finite
value of the quasiparticle density of states (DOS) at zero bias. These facts
could be explained, at least partially, by assuming a {\it d}-wave pair
state which has nodes within the superconducting gap. On the other hand, in
the last years, a few attempts have been made to fit some of the high-$T_c$
superconducting properties by using the Migdal-Eliashberg theory \cite
{refe1,refe2,refe3}, but most of the works concerned only the solution in 
{\it s}-wave pair symmetry. In particular, we recently used the
superconducting DOS of optimally-doped Bi 2212 single crystals with $%
T_c\simeq $ 93 K, determined in break-junction tunnel experiments, in order
to reproducibly determine the electron-boson spectral function of Bi 2212 by
inversion of the {\it s}-wave Eliashberg equations \cite{refe5,refe6}.

In the present paper, we extend the real-axis direct solution of the
Eliashberg equations for strong electron-boson coupling to the {\it d}-wave
case and discuss the main features of the solution as function of
temperature and boson coupling strength. In the framework of this model, we
are able to obtain a very good agreement between our experimental
break-junction tunneling data and the theoretical results with regard to
both the DOS shape in all the energy range and the critical temperature.

Our numerical analysis starts from the well-known generalized Eliashberg
equations for the order parameter $\Delta (\omega ,{\bf k})$ and the
renormalization function $Z(\omega ,{\bf k})$. They have kernels that depend
on the retarded interaction $\alpha ^2(\Omega ,{\bf k},{\bf k}^{\prime
})F(\Omega )$, the coulomb interaction $\mu ^{*}\left( {\bf k},{\bf k}%
^{\prime }\right) $, and the effective band $\varepsilon _{{\bf k}}$ \cite
{refe13,refe14,refe15,refe17}. For simplicity we suppose that ${\bf k}$ and $%
{\bf k}^{\prime }$ lie in the CuO$_2$ plane ($ab$ plane) and we neglect the
relatively small band dispersion and the gap in $c$ direction. Therefore we
use a single band approximation where the Fermi line is nearly a circle
while $\phi $ and $\phi ^{\prime }$ are the azimuthal angles of ${\bf k}$
and ${\bf k}^{\prime }$ in the $ab$ plane, respectively. We thus expand $%
\alpha ^2(\Omega ,\phi ,\phi ^{\prime })F(\Omega )$ and $\mu ^{*}(\phi ,\phi
^{\prime })$ in terms of basis functions $\psi _i\left( \phi \right) $ where
the first few functions of lowest order are:

$\psi _0\left( \phi \right) =1$; $\psi _1\left( \phi \right) =\sqrt{2}\cos
\left( 2\phi \right) $; $\psi _2\left( \phi \right) =\sqrt{2}\sin \left(
2\phi \right) $;

In the present paper we study the simplest $d$-wave electron-boson model
interaction expressed by the following separable expressions

\begin{equation}  \label{1}
\alpha ^2(\Omega ,\phi ,\phi ^{\prime })F(\Omega )=\alpha _{00}^2F(\Omega
)+\alpha _{11}^2F(\Omega )\cdot \psi _1\left( \phi \right) \psi _1\left(
\phi ^{\prime }\right)  \eqnum{1}
\end{equation}

\begin{equation}
\mu ^{*}(\phi ,\phi ^{\prime })=\mu _{00}^{*}+\mu _{11}^{*}\cdot \psi
_1\left( \phi \right) \psi _1\left( \phi ^{\prime }\right)  \eqnum{2}
\label{2}
\end{equation}
and we search for a $s+id$ solution given by

\begin{center}
$\Delta (\omega ,\phi )=\Delta _s(\omega )+\Delta _d(\omega )\cdot \psi
_1\left( \phi \right) $

$Z(\omega ,\phi )=Z_s(\omega )+Z_d(\omega )\cdot \psi _1\left( \phi \right) $%
.
\end{center}

With these assumptions the Eliashberg equations in presence of impurities 
\cite{refe18} become:

\begin{eqnarray}
\Delta _s(\omega )Z_s(\omega ) &=&\int_0^{+\infty }P_s(\omega ^{\prime
})\cdot \left[ K_{00+}(\omega ,\omega ^{\prime })-\mu _{00}^{*}\cdot \Theta
\left( \omega _c-\omega ^{\prime }\right) \right] d\omega ^{\prime }+ 
\nonumber  \label{3} \\
&&i\pi \Gamma \cdot P_s(\omega )/\left[ C^2+P_s(\omega )+N_s(\omega )\right]
\eqnum{3}  \label{3}
\end{eqnarray}

\begin{equation}  \label{4}
\Delta _d(\omega )Z_s(\omega )=\int_0^{+\infty }P_d(\omega ^{\prime })\cdot
\left[ K_{11+}(\omega ,\omega ^{\prime })-\mu _{11}^{*}\cdot \Theta \left(
\omega _c-\omega ^{\prime }\right) \right] d\omega ^{\prime }  \eqnum{4}
\end{equation}

\begin{eqnarray}
\left[ 1-Z_s(\omega )\right] \cdot \omega &=&\int_0^{+\infty }N_s(\omega
^{\prime })\cdot K_{00-}(\omega ,\omega ^{\prime })d\omega ^{\prime }+ 
\nonumber  \label{5} \\
&&i\pi \Gamma \cdot N_s(\omega )/\left[ C^2+P_s(\omega )+N_s(\omega )\right]
\eqnum{5}  \label{5}
\end{eqnarray}
where $\Gamma =n_I/N(0)\pi ^2$, $C=cotg(\delta _0)$, $n_I$ is the impurity
concentration, $N(0)$ is the value of the normal DOS at the Fermi energy and 
$\delta _0$ is the scattering phase shift. We have not written the equation
for $Z_d(\omega )$ that is homogeneous. In the weak-coupling case its only
solution is $Z_d(\omega )\equiv 0,$ while in the strong-coupling one, in
principle, there is a possibility that a nonzero solution exists above some
coupling-strength threshold. In the present work, we assume that the stable
solution corresponds to $Z_d(\omega )\equiv 0$ for all the couplings and we
do not consider the rather exotic possibility of $Z_d(\omega )\neq 0$ \cite
{refe14,refe15}. In eqs. (3)-(5), $\omega _c$ is a cut-off energy and we
have 
\begin{eqnarray*}
K_{ii\pm }(\omega ,\omega ^{\prime }) &=&\int_0^{+\infty }d\Omega \alpha
_{ii}^2(\Omega )F(\Omega )\cdot \\
&&\ \left[ \frac{f\left( -\omega ^{\prime }\right) +n\left( \Omega \right) }{%
\Omega +\omega ^{\prime }+\omega +i\cdot \delta }\pm \frac{f\left( -\omega
^{\prime }\right) +n\left( \Omega \right) }{\Omega +\omega ^{\prime }-\omega
-i\cdot \delta }\mp \frac{f\left( \omega ^{\prime }\right) +n\left( \Omega
\right) }{\Omega -\omega ^{\prime }+\omega +i\cdot \delta }\mp \frac{f\left(
\omega ^{\prime }\right) +n\left( \Omega \right) }{\Omega -\omega ^{\prime
}-\omega -i\cdot \delta }\right]
\end{eqnarray*}
where $f\left( \omega ^{\prime }\right) $ and $n\left( \Omega \right) $ are
the Fermi and Bose functions, respectively, and furthermore we know that

\[
P_s(\omega )=\left( 1/2\pi \right) \int_0^{2\pi }Real\left( \Delta _s(\omega
)/\sqrt{\omega ^2-\Delta _s^2(\omega )-\left[ \Delta _d(\omega )\cdot \psi
_1\left( \phi \right) \right] ^2}\right) d\phi 
\]
\[
P_d(\omega )=\left( 1/2\pi \right) \int_0^{2\pi }Real\left( \Delta _d(\omega
)\cdot \psi _1\left( \phi \right) \psi _1\left( \phi \right) /\sqrt{\omega
^2-\Delta _s^2(\omega )-\left[ \Delta _d(\omega )\cdot \psi _1\left( \phi
\right) \right] ^2}\right) d\phi . 
\]
The quasiparticle DOS that is compared to the experimental data is given by

\begin{equation}  \label{6}
N_s(\omega )=\left( 1/2\pi \right) \int_0^{2\pi }Real\left( \omega /\sqrt{%
\omega ^2-\Delta _s^2(\omega )-\left[ \Delta _d(\omega )\cdot \psi _1\left(
\phi \right) \right] ^2}\right) d\phi .  \eqnum{6}
\end{equation}
In the case we want to obtain a $s+d$ solution, we need only to replace the
denominator of $P_d(\omega )$, $P_s(\omega )$ and $N_s(\omega )$ with $\sqrt{%
\omega ^2-\left[ \Delta _s(\omega )+\Delta _d(\omega )\cdot \psi _1\left(
\phi \right) \right] ^2}.$

In our numerical analysis we have put, for simplicity, $\alpha
_{11}^2F(\Omega )$=$g_d\alpha _{00}^2F(\Omega )$ where $g_d$ is a constant 
\cite{refe19}. As a consequence, the electron-boson coupling constants for
the $s$-wave channel and the $d$-wave one are $\lambda _s=2\int_0^{+\infty
}d\Omega \alpha _{00}^2F(\Omega )/\Omega $ and $\lambda _d=\left( 1/\pi
\right) \int_0^{2\pi }d\phi \psi _1^2\left( \phi \right) \int_0^{+\infty
}d\Omega \alpha _{11}^2F(\Omega )/\Omega =g_d\cdot \lambda _s$, respectively$%
.$

We solved the generalized, real-axis, Eliashberg equations (3)-(5) in direct
way by using an iterative procedure that continues until all the real and
imaginary values of the functions $\Delta _s(\omega ),$ $\Delta _d(\omega )$
and $Z_s(\omega )$ at a new iteration show differences less than $1\cdot
10^{-3}$ with respect to the values at the previous iteration. Usually the
convergence occurs after a number of iterations between 10 and 15. By using
this approach we found that $\Delta (\omega ,\phi )$ has either a pure {\it s%
}-wave or a pure {\it d}-wave symmetry depending on the values of the
coupling constants $\lambda _s$ and $\lambda _d$. The values of the
parameter $g_d=\lambda _d/\lambda _s$ determine the stability of the two
types of solution: if $1.5<g_d<5$, there are no problems for an automatic
convergence to the pure {\it d}-wave solution. When $g_d<1.5$, the choice of
the starting values of $\Delta _s(\omega )$ and $\Delta _d(\omega )$
determines the symmetry of the final solution to which the iteration
procedure converges. If $g_d\gg 1$ (i.e. roughly $g_d>7$) the iteration
procedure converges only in the complex-axis formulation.

We can now check if exists a couple of $\lambda _s$ and $\lambda _d$ values
that, together with a proper electron-boson spectral function $\alpha
^2(\Omega ,\phi ,\phi ^{\prime })F(\Omega ),$ can reproduce our tunneling
density of states $N_{ex}(\omega )$ of Bi 2212 in the framework of a pure $d$%
-wave strong-coupling solution. For doing this we compare the $N_{ex}(\omega
)$ to the quantity $N(\omega )=\int_{-\infty }^{+\infty }N_s(E+\omega
)\left| f(E)-f(E+\omega )\right| dE$ where $N_s(\omega )$ is calculated from
eq.(6) by using $\alpha _{00}^2F(\Omega )=\left( \lambda _s/\lambda \right)
\alpha ^2F(\Omega )_{Bi2212}$. As a first-order approach to the $d$-wave
modeling of the tunneling curves, we used the electron-boson spectral
function previously determined by the inverse solution of the {\it s}-wave
Eliashberg equations applied to the same Bi 2212 data as $\alpha ^2F(\Omega
)_{Bi2212}$ (see the inset of Fig. 1) \cite{refe5,refe6}. Of course, $%
\lambda $ is the corresponding coupling constant.

In Figure 1 our tunneling experimental data (open circles) and the best fit
curve at 4 K (solid line) that is obtained for $\lambda _s=2,$ $\mu
_{11}^{*}=0$, $g_d=1.15$ and yields almost the experimental $T_c,$ $%
T_c^{calc}=99$ K, are shown. We want to emphasize the impressive agreement
of the theoretical curve to the experimental one in all the energy range,
including the region at $V<\Delta _M$ (peak of the conductance) where states
inside the gap in all the experimental tunneling curves are present. Taking
into account a small amount of impurities corresponding to $\Gamma =0.3$ meV
and $C\approx 0,$ also the zero-bias conductance is reproduced and $%
T_c^{calc}=96$ K. In the case of $g_d=1.12$ and no impurities, $%
T_c^{calc}=93 $ K, that is exactly the experimental value, but the fit is a
little less good. It is interesting to note that there are different pairs
of $\lambda _s $ and $\lambda _d\ $values with $2\leq \lambda _s\leq 2.5$
and $2\leq \lambda _d\leq 2.25$, all giving $T_c^{calc}=93$ K and producing
a good fit of the experimental data, the best remaining that one shown in
Fig.1. When the $d$-wave component of the Coulomb pseudopotential is
different from zero (the $s$-wave component does not affect the results in
pure $d$-wave symmetry) the above mentioned best-fit parameters give the
experimental $T_c$ value for $\mu _{11}^{*}=0.0125$. The short dash line of
Fig. 1 shows the normalized $dI/dV$ curve obtained at 4 K in this case. It
is practically indistinguishable from the curve at $\mu _{11}^{*}=0$. In
general, the main consequence of a $\mu _{11}^{*}\neq 0$ is the reduction of
the values of $T_c^{calc}$ and $\Delta _M$. In Fig. 1 we also show the
theoretical normalized conductances, obtained by using the parameters of the
best-fit curve, at different temperatures between 60 K (long dash line) and
140 K (tiny short dash one). We can see that the normalized conductance is
practically flat only at $T^{*}\simeq 140$ K $>T_c$: this is a
strong-coupling effect that occurs at $\lambda \geq 2$ and has been briefly
discussed in $s$-wave symmetry \cite{refe20,refe21,refe22,refe23,refe24}.
Here we demonstrate that it also exists in $d$-wave symmetry, is enhanced by
the particular characteristics of Bi 2212 and could be related to the
pseudogap observed in underdoped and overdoped Bi 2212 \cite{refe12}.

In this $d$-wave strong-coupling regime we have observed other unusual facts
already partially predicted in $s$-wave symmetry \cite
{refe20,refe21,refe22,refe23,refe24}, but here enhanced by the specific
properties of Bi 2212. Some of them can be discussed with reference to Fig.
2 where the real and imaginary parts of the renormalization function $%
Z_s(\omega )$ (Fig. 2a) and of the gap function $\overline{\Delta }_d(\omega
)=\sqrt{2}\Delta _d(\omega )$ (Fig. 2b) are shown for $T=4.2,$ $99$ and $140$
K. In particular, one can observe that the function $Real(\overline{\Delta }%
_d(\omega ))=\overline{\Delta }_{d1}(\omega )$ goes to zero very abruptly
for $\omega \rightarrow 0$ and $T\simeq T_c$ (see the short dash curve of
Fig. 2b), but, in general, at that temperature it is considerable different
from zero at higher energies, whereas in the weak and intermediate coupling
regime the whole curve goes to zero in uniform way at the increase of
temperature. In addition, Fig. 1 shows that the normalized conductance is
practically flat at $T>T^{*}$ but, at the same temperature, $\overline{%
\Delta }_d\left( \omega \right) $ is still different from zero, as it is
suggested by dot and short dot curves of Fig. 2b that are calculated at $%
T=140$ K.

In the best fit case represented by the continuous line of Fig. 1 and at $%
T=4 $ K, our theoretical model produces $\overline{\Delta }_0=\sqrt{2}\Delta
\left( \omega =0\right) =21.22$ meV that gives a ratio $2\overline{\Delta }%
_0/k_BT_c^{calc}=4.98,$ but the values of the gap edge $\overline{\Delta }_g=%
\sqrt{2}\Delta _g=\sqrt{2}%
\mathop{\rm real}
\left[ \Delta \left( \omega =\Delta _g(T)\right) \right] $ and the peak $%
\Delta _M$ are rather different: $\overline{\Delta }_g=24.5$ meV and $\Delta
_M=25$ meV. In Fig. 3a we can see the plot of $\overline{\Delta }_0$, $%
\overline{\Delta }_g$ and $\Delta _M$ as function of temperature: the last
one (solid triangles) does not go to zero at $T_c$ and, in addition, it
grows considerably at larger $T$, being thus present also at $T\geq T_c$.
Particularly at $T>T_c/2$ this peak is very different from the gap edge $%
\overline{\Delta }_g$ (solid squares). On the other hand, it can be seen in
Fig. 3a that both $\overline{\Delta }_g$ and $\overline{\Delta }_0$ have a
BCS-like temperature dependency, but with a different $T_c$ of the order of
115 K and 99 K, respectively. This effect is typical of a very
strong-coupling regime while in the weak and intermediate coupling $%
\overline{\Delta }_g$ and $\overline{\Delta }_0$ always coincide.

Another very interesting and unusual phenomenon related to $\overline{\Delta 
}_g$ occurs for high values of $\lambda _s$ \cite{refe24}. When $\lambda
_s\geq 2$, a $T^{**}$ exists for which at $T\geq T^{**}$ (in our case $%
T^{**}=65$ K) there is an energy value $\omega ^{*}(T)$ that verifies the
expression: $real(Z_s(\omega ^{*}(T)))=imag(Z_s(\omega ^{*}(T)))$. In this
case the quasiparticle approximation is no more valid and it is possible to
have more than one solution of the equation $\Delta _g(T)=%
\mathop{\rm real}
\left[ \Delta \left( \omega =\Delta _g(T),T\right) \right] $. In Fig. 3b we
show the temperature dependencies of the two solutions $\overline{\Delta }%
_{g1},$ $\overline{\Delta }_{g2}$ of the previous equation, obtained for the
spectral function of Fig. 1 and the parameters of the solid line curve of
the same figure, together with the $\omega ^{*}(T)$ curve and its
approximate value $\omega _{th}^{*}(T)$ given by 
\begin{equation}
\omega _{th}^{*}(T)\simeq \left[ 2\pi /(1+\lambda _s)\right] \int_0^{+\infty
}\alpha _{00}^2F(\Omega )\cdot \left[ f\left( \Omega \right) +n\left( \Omega
\right) \right] d\Omega .  \eqnum{7}  \label{7}
\end{equation}
This effect is also present in pure $s$-wave symmetry and is strongly
enhanced by the particular shape of the spectral function of Bi 2212. In
conventional strong-coupling superconductors the range of temperature where
two solutions of the gap equation are present is of the order of $10^{-2}$ K 
\cite{refe24}, while here it is of the order of $50$ K. Particularly in the
present situation, this fact suggests that the gap edge is no more the most
relevant physical quantity$.$ As a consequence, the gap, experimentally
determined from different measurements, is not the gap edge $\Delta _g$ but
an average over the frequency of the function $\Delta \left( \omega
,T\right) $ with a weight factor depending on the property being measured 
\cite{refe24}. This fact could have important consequences in the
interpretation of the experimental data in Bi 2212.

In summary, we have shown that the simplest approach to the pure $d$-wave
solution of the equations for the retarded strong electron-boson interaction
reproduces very well and in the whole energy range the normalized
conductance of optimally-doped Bi 2212, recently determined in our
break-junction tunneling experiments \cite{refe5,refe6}. The addition of a
small amount of impurity scattering in the unitary limit accounts also for
the small but finite value of the normalized conductance at zero bias \cite
{refe27}. In the best-fit cases, the calculated critical temperatures differ
no more than 6\% from the experimental ones. The temperature dependency of
the normalized conductance, calculated for the best-fit values of the
electron-boson coupling constants, shows non-flat curves at $T>T_c$ up to $%
T^{*}\approx 140$ K. This numerical result seems to suggest the presence of
a strong-coupling ''pseudogap'' even in the DOS of optimally-doped Bi 2212
samples where, up to now, it has never been clearly observed. The evolution
of the present work will be its extension to the non half-filling case, in
order to compare the $d$-wave solution of the Eliashberg equations to the
DOS of underdoped and overdoped Bi 2212, and the comparison with other
experimental data. Nevertheless, we believe the generality of the Eliashberg
formalism and its independence of a specific assumption for the microscopic
coupling mechanism gives the present results a hardly questionable validity.

{\bf FIGURE\ CAPTIONS}\\

\noindent Fig. 1 The theoretical normalized conductances for $\mu
_{11}^{*}=0 $ at various temperature values: $T=4,$ $60,$ $100,$ $120,$ $140$
K and for $\mu _{11}^{*}=0.0125$ at $T=4$ K, calculated with $\lambda _s=2$
and $\lambda _d=2.3$. The optimally-doped Bi 2212 break-junction
experimental data at $T=4.2$ K are shown as open circles. In the inset, the $%
\alpha ^2F\left( \Omega \right) _{Bi2212}$ used in the fit of experimental
data is shown.\\

\noindent Fig. 2 (a) The calculated values of $ReZ_s(\omega )=Z_{s1}(\omega
) $ and $ImZ_s(\omega )=Z_{s2}(\omega )$ at three different temperatures $%
T=4.2 $, $99$, $140$ K; (b) the same as in (a) but for $Re\overline{\Delta }%
_d(\omega )=\overline{\Delta }_{d1}(\omega )$ and $Im\overline{\Delta }%
_d(\omega )=\overline{\Delta }_{d2}(\omega )$.\\

\noindent  Fig. 3 (a) Calculated values of $\Delta _M(T)$ (peak of the DOS), 
$\overline{\Delta }(\omega =0,T)$ and $\overline{\Delta }_g(T)$ (gap edge)
as function of temperature. The solid and dash lines represent the BCS
dependence of the gap for $T_c=115$ K and $T_c=99$ K, respectively; (b) the
two solutions of the gap equation $\overline{\Delta }_{g1}$ and $\overline{%
\Delta }_{g2}$, together with $\omega ^{*}(T)$ and its approximate
analytical expression (short dash line) as function of temperature.\\.

\end{document}